\documentclass[conference]{IEEEconf}
\input epsf
\usepackage{balance} 
\usepackage{stfloats}
\usepackage{cite}
\usepackage{amsmath,amssymb,amsfonts}
\usepackage{algorithmic}
\usepackage{graphicx}
\usepackage[table,xcdraw]{xcolor}
\usepackage{textcomp}
\usepackage[numbers]{natbib}
\usepackage{todonotes}
\usepackage{tcolorbox}
\usepackage{multirow}
\usepackage[hidelinks]{hyperref}
\usepackage{url}
\usepackage{pifont}
\usepackage{tabularx} 
\def\BibTeX{{\rm B\kern-.05em{\sc i\kern-.025em b}\kern-.08em
    T\kern-.1667em\lower.7ex\hbox{E}\kern-.125emX}}

\renewcommand\thesection{\arabic{section}} 

\renewcommand\thesubsectiondis{\thesection.\arabic{subsection}}

\renewcommand\thesubsubsectiondis{\thesubsectiondis.\arabic{subsubsection}}

\begin{document}

\author{Jesse Nyyssölä$^{*}$, Mika Mäntylä\\
	\normalsize University of Helsinki, Helsinki, Finland\\
	\normalsize jesse.nyyssola@helsinki.fi, mika.mantyla@helsinki.fi\\
	\normalsize *corresponding author
}

\title{\textbf{\Large Speed and Performance of Parserless and Unsupervised Anomaly Detection Methods on Software Logs\\}}

\maketitle

\noindent \begin{abstract}
Software log analysis can be laborious and time consuming. Time and labeled data are usually lacking in industrial settings. This paper studies unsupervised and time efficient methods for anomaly detection. We study two custom and two established models. The custom models are: an OOV (Out-Of-Vocabulary) detector, which counts the terms in the test data that are not present in the training data, and the Rarity Model (RM), which calculates a rarity score for terms based on their infrequency. The established models are KMeans and Isolation Forest. The models are evaluated on four public datasets (BGL, Thunderbird, Hadoop, HDFS) with three different representation techniques for the log messages (Words, character Trigrams, Parsed events). For training, we used both normal-only data, which is free of all anomalies, and unfiltered data, which contains both normal and anomalous instances. We used primarily the AUC-ROC metric for evaluation due to challenges in setting a threshold but we also include F1-scores for further insight.  
Different configurations are advised based on specific requirements. When training data is unfiltered, includes both normal and anomalous instances, the most effective combination is the Isolation Forest with event representation, achieving an AUC-ROC of 0.829. If it's possible to create a normal-only training dataset, combining the Out-Of-Vocabulary (OOV) detector with trigram representation yields the highest AUC-ROC of 0.846. For speed considerations, the OOV detector is optimal for filtered data, while the Rarity Model is the best choice for unfiltered data.

\end{abstract}
\IEEEoverridecommandlockouts
\vspace{1.5ex}

\begin{keywords}
\itshape anomaly detection; novelty detection, software logs; log event; log analysis; machine learning; unsupervised learning
\end{keywords}

\IEEEpeerreviewmaketitle

\section{Introduction}

\nocite{yu2023deep, logrepresentation, landauer2023deep, pinpoint, cao2022higher, confmasked}

Software anomaly detection is gravitating towards more complex algorithms, despite evidence suggesting that computationally intensive deep-learning approaches may not offer significant benefits \cite{yu2023deep}-\cite{confmasked}. 
This study aims to examine the anomaly detection approaches that are both fast and require little preliminary effort by the log owner. Primarily this means that we are using unsupervised methods that do not require labelled anomalies in the training data. Labeling large log files is a laborious endeavor and in industrial settings the labels are rarely available \cite{expreport}. 

In this paper, we examine methods that do not require labeled data, focusing on two scenarios: Unfiltered and Normal-only. In the Unfiltered scenario, we assume the log owner lacks information about the logs, and the model is trained on completely unfiltered data, containing both anomalies and normal data. In the Normal-only scenario, although the log owner remains unaware of the type or location of anomalies, they can isolate clean log data, devoid of anomalies, for training purposes. Research on the difference between Unfiltered and Normal-only has been very scarce despite its fundamental impact on detection task.
In the Normal-only scenario, anomaly detection will be based on the dissimilarity of the log messages in the test and training data. This approach has been demonstrated by previous work numerous times, e.g. \cite{pinpoint},\cite{deeplog}, and is sometimes referred to as "novelty detection" \cite{language_novelty}.

Furthermore, we examine the effect of log representation to see whether representations with lower computational cost can prove effective. 
A recent study \cite{logrepresentation} conducted a thorough analysis on log representations used in anomaly detection. The authors applied several classical and semantic-based representations on traditional and deep-learning models. One of the key findings was that the traditional models performed better with classical representations which is a major motivation for our work. A widely used technique for log representation is to use a clustering, also known as log parsing, algorithm, to find out the static part of a log message called the template and use that (or its ID) to represent the logline \cite{literaturereview}. As this step is computationally expensive we will use more straight forward approaches for representing the log messages such as word-based splitting and character trigram generation to assess their effect on the performance of the selected models. Based on these insights, we formulated the following three research questions:

\begin{itemize}
    \item \textbf{RQ1}: How good are the unsupervised models for anomaly detection as measured with AUC-ROC, F1-score and time?
    \item \textbf{RQ2}: How does log representation affect the results: Words vs. trigrams vs. event ID?
    \item \textbf{RQ3}: Do the models perform better after filtering out anomalies from the training data? 
\end{itemize}

This study will proceed by presenting the most relevant related work. In the Research Methods chapter we describe the preprocessing, datasets, metrics and models used in the experiments. The Results chapter presents the results as divided by the three chosen metrics (AUC-ROC, F1-score and time). The implications of the results along with future research directions and limitations are discussed in the Discussion chapter. After that we present the conclusion of the study. 

\section{Related work}

A systematic literature review \cite{literaturereview} provides a general overview on the classification and approaches used in log based anomaly detection. It provides tools to position one's study based on preprocessing type, metrics, datasets and the various detection methods. The review 
shows that majority of the recent studies are based on computationally heavy approaches such as deep learning or utilize supervised models.  Another literature review \cite{reviewparsing} notes the lack of real world use cases and reports Loglizer\footnote{https://github.com/logpai/loglizer} as the only available toolkit for log anomaly analysis. 
The results from the Loglizer toolkit have been reported in a previous study \cite{expreport}. 
In the study, Invariant Mining reached the best results with F1-score of 0.91 on both HDFS and BGL datasets. Regarding unsupervised models, the authors note challenges of setting and testing out various thresholds \cite{expreport}. The Loglizer paper notes that unsupervised models are also much more time consuming than supervised ones. For example, 1.5 million lines on HDFS took over 1000 seconds with Invariant Mining. 

LogAnomaly \cite{loganomaly} utilizes an unsupervised LSTM model that combines word and template vectors and manages to reach F1-scores of 0.96 (BGL) and 0.95 (HDFS). However, it has been demonstrated that deep learning-based models are sensitive to hyperparameter tuning \cite{empiricalml} and they are also computationally expensive. 

AutoLog \cite{autolog} focuses on training the model with non-anomalous data and predicting anomalies based on reconstruction error. Their model uses a deep autoencoder and manages to reach F1-scores of 0.95 on BGL and 0.97 on HDFS. Additionally, they compare their model with unsupervised models such as the IsolationForest which reached worse scores of 0.70 on BGL and 0.61 on HDFS. 

Another study on unsupervised models \cite{comparitive_unsup} showcases various models and evaluates them using the Area Under Curve metric. Their results suggest the k-nearest neighbour, Local Outlier Factor and  Histogram-based Outlier Score algorithms. However, these are generalized results based on datasets from multiple domains.

NeuralLog \cite{withoutparsing} introduces an anomaly detection approach that works without parsing. By doing this, it aims to solve the issue of Out-of-vocabulary (OOV) words and semantic misunderstandings. NeuralLog achieves great results on several datasets but it employs a BERT encoder and transformer-based classification \cite{withoutparsing} which are relatively time consuming approaches. 

\section{Research methods}
This chapter describes the preprocessing, datasets, metrics and models used in the study. Our environment consists of a virtual machine that has Intel Core Processor (Broadwell, IBRS) with 28 cores at 2.4 Ghz and 224 GB of memory. This study was conducted on a modified version of the LogLead\footnote{https://github.com/EvoTestOps/LogLead} tool that is designed to streamline the process from loading the log file to anomaly detection \cite{loglead}. The replication package is publicly available on GitHub\footnote{https://github.com/jnyyssol/LL-mod-unsupervised}. 

\subsection{Preprocessing}
The data is loaded directly from raw log files to Polars \cite{Vink2023Polars} dataframes. 
Next, the log messages are normalized to reduce the effect of parameters (non-static parts of the message). However, due to placing strong emphasis on performance time we only perform three operations with fast Polars-native string operations: 
\begin{enumerate}
    \item Convert all letters to lowercase, 
    \item Replace every numeral with the digit '0',
    \item Reduce any consecutive series of zeros to a single '0'.
\end{enumerate}
For example, the string "Time 12:34:56" would be converted to "time 0:0:0". When it comes to words and trigrams, normalization will make the resulting vocabulary smaller. While the primary reason for this is to improve the accuracy due to minimizing the effect of parameters, smaller vocabulary also makes the model faster. 

The normalized log messages are transformed into event IDs or lists consisting of trigrams or words. Event IDs are created with the Drain parser \cite{drain}. 
By "trigrams" we refer to character trigrams, that is, three consecutive characters in the log message. For example, the first three trigrams of the message "Stopping..." would be "Sto", "top" and "opp".  
Words are extracted with a Polars native string split by using whitespace as the separator. An example of these representations along with their raw and normalized log messages are shown in Table~\ref{tab:preprocess}.

\begin{table*}[ht]
\caption{Example log line from the preprocessed BGL dataset}
\label{tab:preprocess}
\centering
\begin{tabular}{|p{0.15\textwidth}|p{0.2\textwidth}|p{0.2\textwidth}|p{0.2\textwidth}|p{0.1\textwidth}|}
\cline{3-5}
\multicolumn{2}{c|}{} & \multicolumn{3}{c|}{\textbf{Representations}} \\ 
\hline
\textbf{Raw log message} & \textbf{Normalized message} & \textbf{Words} & \textbf{Trigrams} & \textbf{Event ID} \\ 
\hline
'4 ddr error(s) detected and corrected on rank 0, symbol 11 over 20609 seconds' & '0 ddr error(s) detected and corrected on rank 0, symbol 0 over 0 seconds' & [ '0', 'ddr', 'error(s)', 'detected', 'and', 'corrected', 'on', 'rank', '0', 'symbol', '0', 'over', '0', 'seconds'] & ['0 d', 'dd', 'ddr', 'dr ', 'r e', 'er ', 'err', 'rro', 'ror', 'or(', '(r(s', '(s)', 's) ', ') d', ' d ', 'de', 'det', 'ete', 'tec', 'ect', ...] & 'e10' \\ 
\hline
\end{tabular}

\end{table*}

The data is then divided into training and test sets based on proportion defined in Table~\ref{datasummary}. From the training set, we create two instances: One with full data and another where anomalies are filtered out. These will be referred as the Unfiltered and Normal-only setups respectively. In an industrial setting, it is not feasible to filter out anomalies. Instead, the idea is that there would be separate logs that are gathered during the system's normal operation that should work as the training data. Hence, the normal-only setup should not been seen as an approach that requires labels as that would undermine it being "unsupervised" but it does require a certain level of curation from the log owner. 

The final step of the preprocessing is to turn the training and test sets into matrices with a 
vectorizer. For most models we use the TfidfVectorizer \cite{scikit-learn} which is a common technique for feature extraction that adds weight to the most important tokens. However, for the Out-of-vocabulary Detector we utilize absolute counts so we use the CountVectorizer.

\subsection{Datasets}
The datasets used in the study are BGL \cite{bgl}, Thunderbird \cite{bgl}, HDFS \cite{hdfs} and Hadoop \cite{hadoop}, which are easily accessible via LogHub \cite{loghub}. Table~\ref{datasummary} shows the summary of the datasets. BGL and Thunderbird are missing the rows for number of sequences as they do not have them and they are labeled at line level. Similarly, the number of normal log lines cannot be determined for HDFS and Hadoop because they are labeled on the sequence level. 

The most significant difference with the sequence and event based datasets for anomaly detection is that when using event ID as the log representation, the sequence based datasets have multiple IDs to influence the prediction while event based ones only have one. For word- and trigram-representations, the sequence based datasets are handled by flattening the lists of lists into a single list with a single anomaly label.

For the event-based datasets, we could create artificial sequences such as previous work has done \cite{confmasked}, \cite{howfar} which would give us a unified way to handle all datasets. However, we think that anomaly detection should be done at the most detailed level possible which means using events instead of sequences if we can. This is especially true if for datasets like BGL and Thunderbird that do not have an obvious way of creating the sequences such as a sequence ID.

We generally wanted to use the same settings for all datasets which means using all data and a small split for training. DeepLog has demostrated that for HDFS a less than 1\% split for training is sufficient \cite{deeplog}. As this paper does not aim to study split proportion, we chose a higher value of 5\% for everything except Hadoop (which has 50\% due to a very small number of sequences). This amount should provide a good coverage of the log lines present in the normal data. Due to memory constraints we had to represent Thunderbird with a randomly chosen 10\% sample of the whole data. The sample is still over 20 million lines. For Thunderbird, the 5\% training data is taken from this sample, so in effect, the training data is just 0.5\% of the total data. 

\subsection{Metrics}

In the study we will analyze model performance with AUC-ROC and F1-score. 
AUC-ROC \cite{aucroc} is a combination of the abbreviations Area Under Curve (AUC) and Receiver Operating Characteristic (ROC). ROC curve is a plot of the true positive rate against the false positive rate across a range of thresholds. With AUC we refer to the area under this curve. This area goes from 0 to 1 and in practice it means the probability that a randomly selected positive sample is ranked higher than a negative one \cite{aucroc}. F1-score is the most common metric in anomaly detection that measures the harmonic mean of precision and recall.

The main reason for emphasizing AUC-ROC in this study is that it is not dependent on a threshold. With supervised models you could assess the model specific anomaly scores in relation to the labels and adjust the threshold accordingly. In unsupervised models this is not possible which leads to subjective threshold selection.

As realistic threshold selection (i.e., without labels) for the F1-score is out of scope of this paper, we will utilize labels in a Bayesian optimization loop that finds the best threshold over a maximum of 20 iterations. We emphasize that this is not possible without labels, and as such, the F1-scores serve as an example of what could be possible. However, from the results we can gain insight on what would be a good way to set the threshold in the future. This is discussed further in the Discussion section. 

The model execution time is reported as the total of training and predicting with a given model. Preprocessing times will be reported separately as it consists of several stages, but only one representation type is required by each model. 
 
\begin{table*}
\caption{Data summary}
\centering
\begin{tabularx}{0.8\textwidth}{
  >{\hsize=1.35\hsize}X
  >{\hsize=0.85\hsize}X
  >{\hsize=0.75\hsize}X
  >{\hsize=0.75\hsize}X
  >{\hsize=1.3\hsize}X}
\hline
\textbf{Dataset} & \textbf{HDFS} & \textbf{BGL} & \textbf{Hadoop} & \textbf{Thunderbird} \\
\hline
Log lines & 11,175,629 & 4,747,963 & 394,308 & 211,212,174 \\
Normal sequences  & 558,223 & - & 11 & - \\
Anomalous sequences  & 16,838 & - & 43 & - \\
Normal lines & -  & 4,399,503 & - & 193,744,715  \\
Anomalous lines & - & 348,460 & - &  17,467,459 \\
Data used & 100\% & 100\% & 100\% & 10\%  \\
Train portion & 5\% & 5\% & 50\% & 5\% \\
\hline
\end{tabularx}
\label{datasummary}
\end{table*}

\subsection{Models}

We did initial tests on several models but some well-known unsupervised models were found to be too slow, such as the Local Outlier Factor and One-class SVM. Additionally, we created two simple custom models, Out-of-vocabulary Detector and a Rarity Model, as we suspected they would be competitive against the established models at least regarding execution speed. All of the models work by producing a numeric value that correspond to how anomalous a given event or sequence is. As the models work identically with all three log representations (words, trigrams, event IDs), in this chapter we refer to them all collectively as "terms".

\subsubsection{Out-of-vocabulary Detector}
We started to design the Out-of-vocabulary Detector (OOVD) as a tool for aiding other models. However, when training the model with only normal data OOV terms are great candidates for being anomalies as the task of anomaly detection turns into novelty detection \cite{language_novelty}.  As such the OOVD by itself can produce good results on normal-only datasets. 

It is important to note that the OOVD can not work if there are anomalies in the training data vocabulary so it will only be used with the Normal-only training setup. Other models can work with the Unfiltered setup because they determine the anomalousness of a term is by its relation to other terms, but OOVD only checks whether it belongs to a set. Therefore, if the set is created randomly the results will be random as well.

Counting OOV terms in millions of lists individually would be a slow operation so we utilize a strategy that takes advantage of the way OOV terms are handled by vectorizers. When transforming the test data with CountVectorizer that was fitted with training data, the OOV terms will automatically get the value 0. Then we can simply subtract the number of in-vocabulary terms calculated by the CountVectorizer from the total number of terms to get the number of OOV terms. Working with total counts as such can be hundreds of times faster than matching the words individually in a for-loop.

\subsubsection{Isolation Forest}
One model implemented with Sklearn \cite{scikit-learn} is Isolation Forest (IF). It is based on random isolation of features and splitting between maximum and minimum values of selected feature \cite{scikit-learn}. Due to their nature, anomalies are easier to separate than normal data. Based on that, it can be determined that the number of partitions it takes to isolate a data point corresponds with how likely it is to be normal. 

\subsubsection{KMeans}
The second model implemented with Sklearn is the KMeans clustering algorithm \cite{scikit-learn}. It works by starting with random centroids and assigning data points to the closest one. As data is added, the centroids are updated based on the mean of all the data points in that centroid. The trained model should have clearly defined clusters around the centroids. For prediction, we give the anomaly score as the minimum distance from the nearest centroid.

\subsubsection{Rarity Model}
We implemented a simple class based on negative logarithm that we called the Rarity Model (RM). In essence, negative logarithm measures the relative infrequency of individual terms in the data that increases based on their rarity. While utilizing negative logarithm in such a way is not novel, our aim was to create a very efficient baseline model. For implementation, the rarity scores are first precalculated for each term in the training data and assigned into a vector. Next, the dot product of the precalculated vector and the sparse matrix generated by the Tf-idf vectorizer from the test data is calculated. The total rarity score for the log message will be the dot product divided by the total number of terms in the message. Given its simple design and reliance on pre-built matrix multiplication routines, the RM is anticipated to function with remarkable speed.

\section{Results}
This chapter presents the results in three parts according to the metrics under study.
\subsection{AUC-ROC results}

\begin{table}
\centering
\caption{AUC with Unfiltered training data}
\label{tab:auc_uf}
\begin{tabularx}{\columnwidth}{
  >{\hsize=1.3\hsize}X
  >{\hsize=1.1\hsize}X
  >{\hsize=0.9\hsize}X
  >{\hsize=0.9\hsize}X
  >{\hsize=1.0\hsize}X
  >{\hsize=0.9\hsize}X
  >{\hsize=0.9\hsize}X}

\hline
\textbf{Log rep.} & \textbf{Model} & \textbf{BGL} & \textbf{Tb} & \textbf{Hadoop} & \textbf{HDFS} & \textbf{Avg} \\
\hline
\textbf{Words} & IF & 0.672 & \textbf{0.783} & \textbf{0.611} & 0.990 & \textbf{0.764} \\
 & KMeans & \textbf{0.722} & 0.293 & 0.579 & 0.971 & 0.641 \\
 & RM & 0.638 & 0.217 & 0.365 & \textbf{0.999} & 0.555 \\
\hline
\textbf{Trigrams} & IF & 0.675 & \textbf{0.814} & \textbf{0.698} & 0.981 & \textbf{0.792} \\
 & KMeans & \textbf{0.741} & 0.166 & 0.611 & 0.956 & 0.619 \\
 & RM & 0.735 & 0.102 & 0.540 & \textbf{0.999} & 0.594 \\
\hline
\textbf{Events} & IF & 0.755 & \textbf{0.693} & \textbf{0.889} & 0.978 & \textbf{0.829} \\
 & KMeans & \textbf{0.766} & 0.187 & 0.847 & \textbf{0.984} & 0.696 \\
 & RM & 0.581 & 0.222 & 0.486 & 0.929 & 0.555 \\
\hline
\end{tabularx}
\end{table}

\begin{table}
\centering
\caption{AUC with Normal-only training data}
\label{tab:auc_f}
\begin{tabularx}{\columnwidth}{
  >{\hsize=1.3\hsize}X
  >{\hsize=1.1\hsize}X
  >{\hsize=0.9\hsize}X
  >{\hsize=0.9\hsize}X
  >{\hsize=1.0\hsize}X
  >{\hsize=0.9\hsize}X
  >{\hsize=0.9\hsize}X}

\hline
\textbf{Log rep.} & \textbf{Model} & \textbf{BGL} & \textbf{Tb} & \textbf{Hadoop} & \textbf{HDFS} & \textbf{Avg} \\
\hline
\textbf{Words} & OOVD & 0.777 & 0.802 & \textbf{0.786} & 0.543 & 0.727 \\
 & IF & 0.482 & 0.433 & 0.611 & \textbf{0.954} & 0.620 \\
 & KMeans & 0.937 & 0.893 & 0.579 & 0.940 & \textbf{0.837} \\
 & RM & \textbf{0.958} & \textbf{0.967} & 0.381 & 0.950 & 0.814 \\
\hline
\textbf{Trigrams} & OOVD & \textbf{0.997} & \textbf{0.997} & \textbf{0.845} & 0.545 & \textbf{0.846} \\
 & IF & 0.605 & 0.530 & \textbf{0.845} & 0.944 & 0.731 \\
 & KMeans & 0.945 & 0.582 & 0.802 & 0.953 & 0.820 \\
 & RM & 0.988 & 0.714 & 0.619 & \textbf{0.982} & 0.826 \\
\hline
\textbf{Events} & OOVD & \textbf{0.995} & \textbf{0.895} & 0.903 & 0.535 & \textbf{0.832} \\
 & IF & 0.004 & 0.104 & \textbf{1.000} & 0.937 & 0.511 \\
 & KMeans & 0.041 & 0.235 & 0.806 & \textbf{0.981} & 0.515 \\
 & RM & 0.405 & 0.387 & 0.278 & 0.840 & 0.477 \\
\hline

\end{tabularx}

\end{table}

\begin{figure*}
    \centering
    \includegraphics[width=\textwidth]{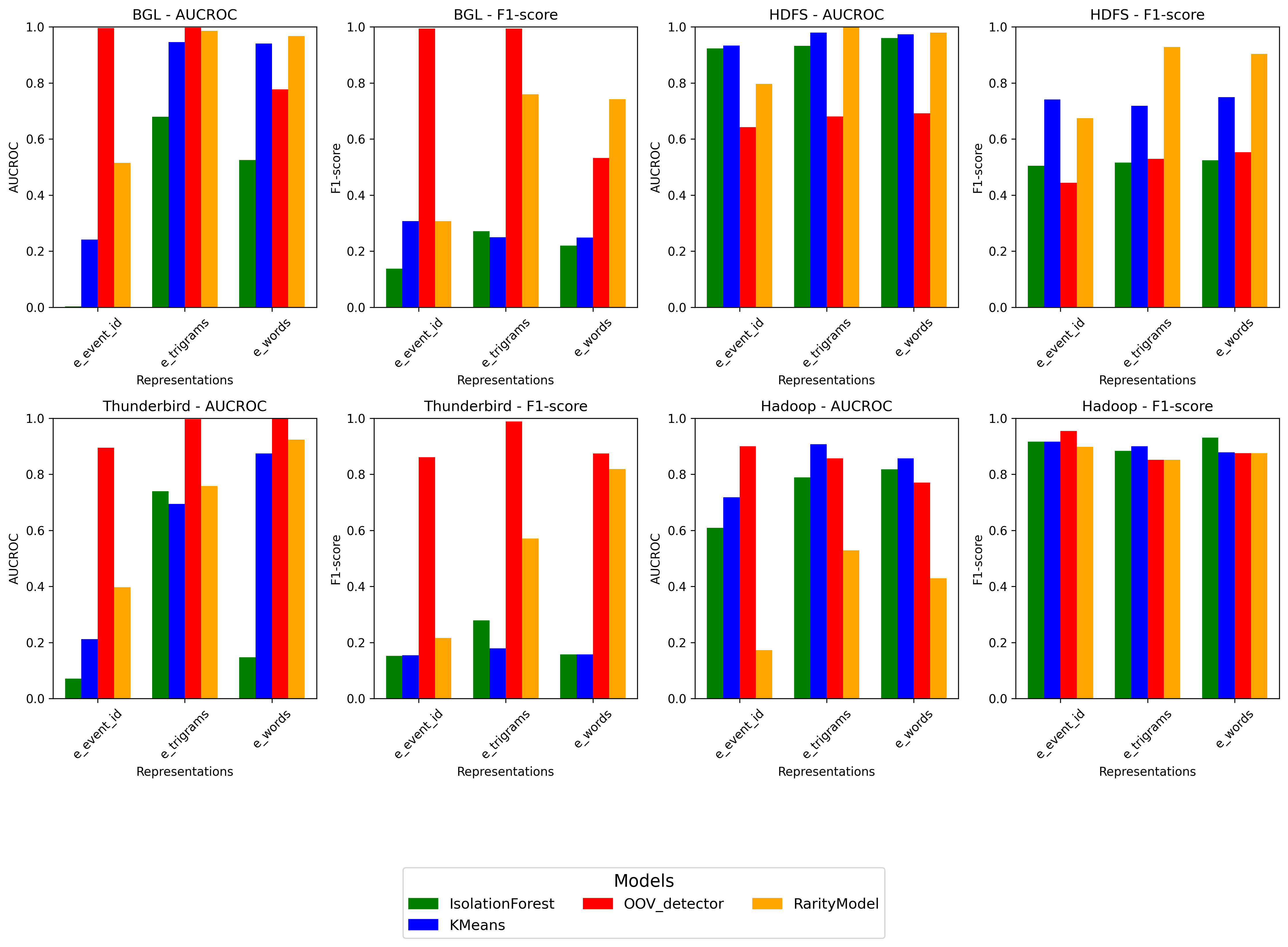}
    \caption{The performance metrics on the Normal-only test setup on all datasets and representations}
    \label{fig:filtered}
\end{figure*}

\begin{figure*}
    \centering
    \includegraphics[width=\textwidth]{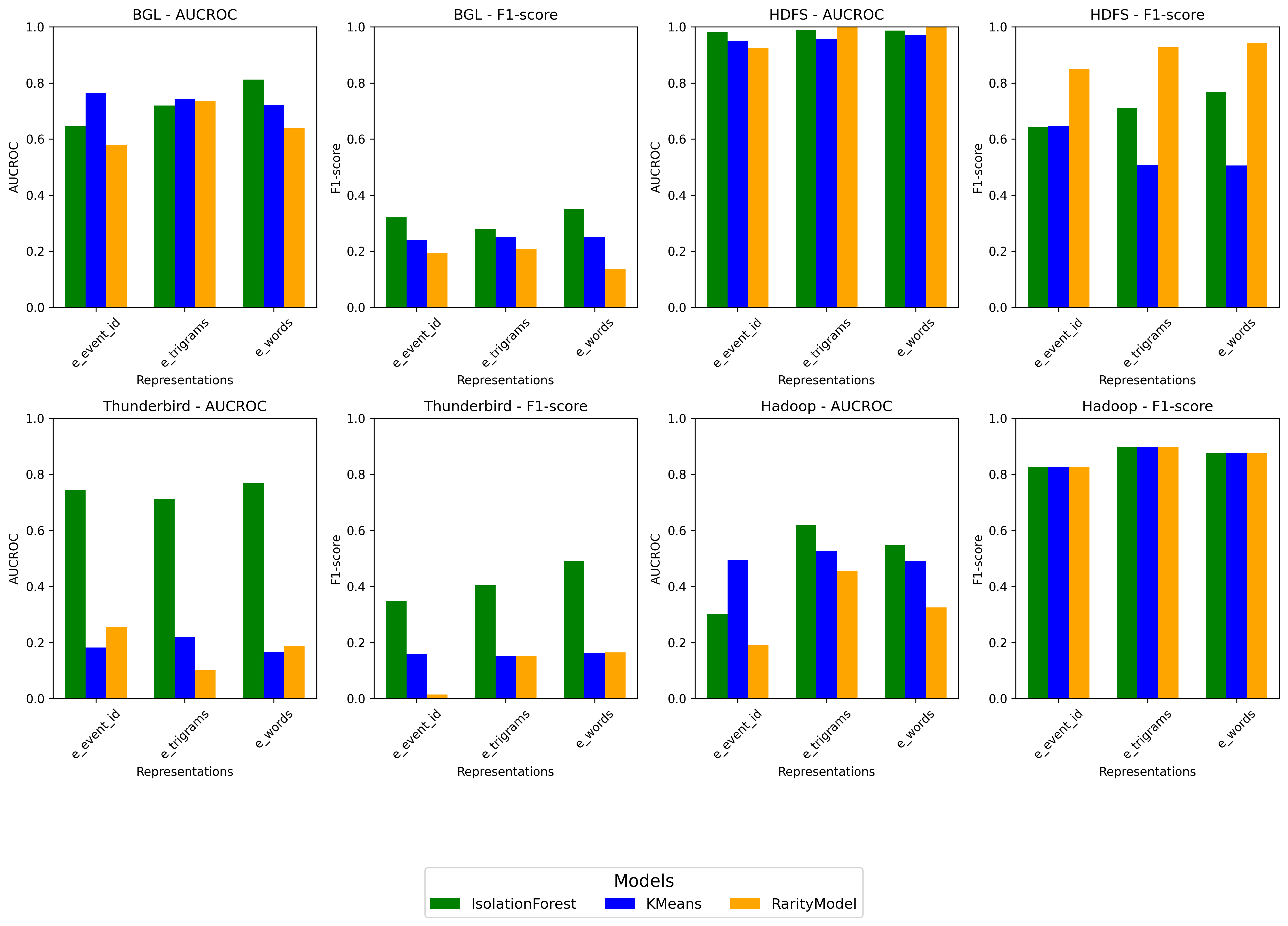}
    \caption{The performance metrics on the Unfiltered test setup on all datasets and representations}
    \label{fig:unfiltered}
\end{figure*}

The results for AUC-ROC are showcased in Table~\ref{tab:auc_uf} and Table~\ref{tab:auc_f}. It is apparent that there is no single model-representation pair that gets the best results on each dataset. IF is a strong candidate for the best model for the Unfiltered scenario where data contains both anomalies and normal observations without labels. KMeans performs similarly to IF on other datasets but fails completely on Thunderbird. RM performs poorly on the Unfiltered data as is to be expected due to the anomalies being in the vocabulary. The reason RM gets a good score with Unfiltered HDFS is that the dataset just has few anomalies in general so when RM sees something rare it will get a higher anomaly score. 

With the Normal-only scenario both KMeans and RM improve the results drastically when using the word- or trigram-representation. It is easy to intuitively understand why this happens on the RM. As the vocabulary only consists of what is normal, the model learns what is normal and deviations from that can be considered abnormal, not just rare. It is interesting that the performance went down with the Normal-only data on both the IF model and the event-representation. Naturally, predicting based on the event ID introduces a lot of variation as BGL and Tb only have one item to predict with due to being event level labeled. Additionally, the modest preprocessing regexes might affect the parsing accuracy. Yet OOVD still managed to get good results with the events on BGL and Tb. This could suggest that the parser works well but other models are not able to handle the OOV events quite well.  

The OOVD did well for everything except HDFS. This is likely due to the nature of the data. If we observe the values that OOVD and RM get, it is usually the case that when one does relatively well, the other does relatively worse. This is caused by the fact that RM (along with IF and KMeans) do not have a way of distinguishing between OOV terms and other zeros in the matrices. Because the vectorizer is fitted on training data and transformed on test data, the terms that appear in the test data but not in the training data (OOV terms) are effectively ignored which is intended functionality for Sklearn vectorizers. In practice, the question that flips the balance is whether the term is never seen or very rarely seen. 

On BGL-trigrams both the RM and OOVD get great results which suggests that the anomalous log messages have both OOV and very rare terms in them. An example of this could be a folder path as the hierarchical structure goes from generic to rare by its nature. The folder example also explains why the OOVD did worse on the word-representation. As the words are not split by slashes, the whole path becomes a long word that is likely to increase the anomaly score. 



 \subsection{F1-score results}

The F1-score reached lower values than AUC-ROC despite using labels for the threshold selection. Figures \ref{fig:filtered} and \ref{fig:unfiltered} illustrate the results for each dataset and representation as bar plots for both F1-score and AUC-ROC. Fig.~\ref{fig:filtered} visualizes well how, like with AUC-ROC, OOVD performs the best with Normal-only training data on F1-score as well. Similarly, Fig.~\ref{fig:unfiltered} shows that IsolationForest remains the top model for the Unfiltered test setup.   

\begin{figure*}
    \centering
    \includegraphics[width=\textwidth]{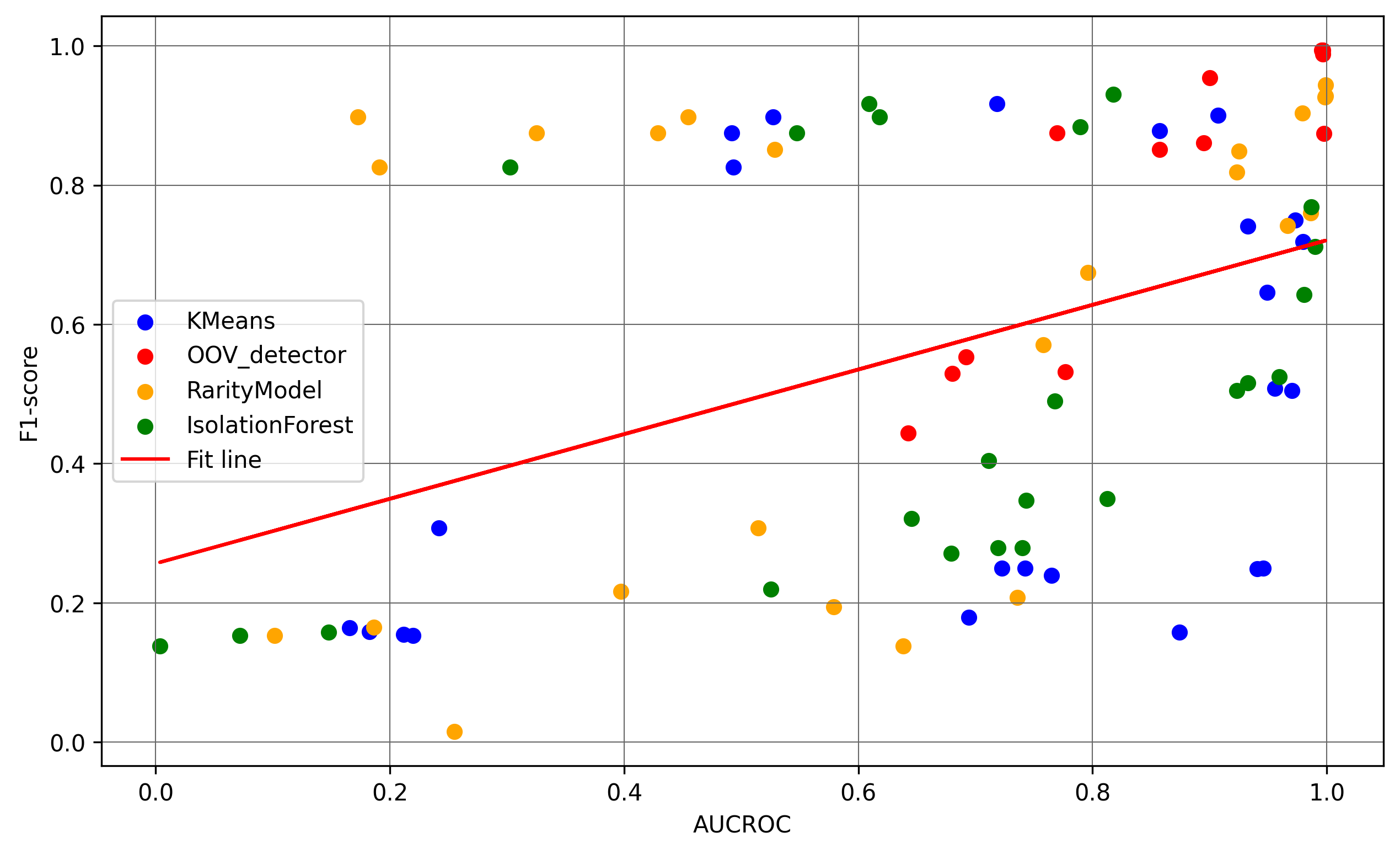}
    \caption{All the results plotted by AUC-ROC and F1-score}
    \label{fig:f1vaucroc}
\end{figure*}

Finally all the scores are summarized in Fig.~\ref{fig:f1vaucroc} along with a fit line that describes the correlation between the AUC-ROC and F1 metrics. The Pearson correlation coefficient is 0.43 with p-value of \( 3.71~\times~10^{-5}\) which indicates statistically significant moderate correlation between the metrics. The correlation is apparent in cases where a model performed well on both metrics. There is also a cluster of cases where both performed poorly. The extent to which the F1-score and the AUC-ROC differ is largely due to the dataset. For example, in the plot the row of high F1-scores at the top are almost all from Hadoop. Similarly, the cluster of high AUC-ROC and low F1-scores are from Thunderbird and BGL.

\subsection{Time results}

\begin{table}
\centering
\caption{Time with filtered training data (seconds)}
\label{tab:time_f}
\begin{tabularx}{\columnwidth}{
  >{\hsize=1.3\hsize}X
  >{\hsize=1.1\hsize}X
  >{\hsize=0.9\hsize}X
  >{\hsize=0.9\hsize}X
  >{\hsize=1.0\hsize}X
  >{\hsize=0.9\hsize}X
  >{\hsize=0.9\hsize}X}

\hline
\textbf{Log rep. } & \textbf{Model} & \textbf{BGL} & \textbf{Tb} & \textbf{Hadoop} & \textbf{HDFS} & \textbf{Avg} \\
\hline
\textbf{Words} & OOVD & \textbf{1.29} & \textbf{5.30} & \textbf{$<$0.01} & 0.20 & \textbf{1.70} \\
 & IF & 34.00 & 185.87 & 0.13 & 9.49 & 57.37 \\
 & KMeans & 2.58 & 13.97 & 0.31 & 0.53 & 4.35 \\
 & RM & 1.62 & 9.27 & \textbf{$<$0.01} & \textbf{0.19} & 2.77 \\
\hline
\textbf{Trigrams} & OOVD & 2.96 & 34.53 & \textbf{$<$0.01} & \textbf{0.36} & 9.46 \\
 & IF & 93.24 & 999.70 & 0.15 & 48.46 & 285.39 \\
 & KMeans & 4.16 & 25.35 & 0.03 & 1.61 & 7.79 \\
 & RM & \textbf{2.02} & \textbf{11.66} & 0.01 & 0.43 & \textbf{3.53} \\
\hline
\textbf{Events} & OOVD & \textbf{0.74} & \textbf{4.31} & \textbf{$<$0.01} & \textbf{0.18} & \textbf{1.31} \\
 & IF & 21.90 & 106.43 & 0.17 & 4.53 & 33.26 \\
 & KMeans & 2.17 & 10.56 & 0.03 & 0.48 & 3.31 \\
 & RM & 1.37 & 7.83 & 0.01 & 0.23 & 2.36 \\
\hline

\end{tabularx}

\end{table}

There were no substantial differences between the run times on the unfiltered and filtered setups. Table~\ref{tab:time_f} show that OOVD and RM are generally the two fastest approaches with KMeans being a little slower. Depending on the data size and use case, IF can become infeasible due to the low speed. For example, with the trigram-representation IF could be 100 times slower than RM on certain datasets. Regarding log representation, the runtime is dependent on the number of items in the data. Therefore, trigrams are the slowest while events are the fastest.  

If we account for pre-processing as shown in Table~\ref{tab:pretimes}, it becomes apparent that the word-representation is the fastest end-to-end approach when starting with a raw log file. Generating trigrams turns out to be a slow process with the standard Python tools. In fact, it was slower than the parser. However, with further optimization, in Polars in particular, it should be possible to achieve speeds that are closer to word creation speed.

\begin{table}
\centering
\caption{Pre-processing times (seconds)}
\label{tab:pretimes}
\begin{tabularx}{\columnwidth}{
  >{\hsize=1.8\hsize}X
  >{\hsize=0.8\hsize}X
  >{\hsize=0.8\hsize}X
  >{\hsize=0.8\hsize}X
  >{\hsize=0.8\hsize}X}
\hline
\textbf{Time to...} & \textbf{Hadoop} & \textbf{BGL} & \textbf{HDFS} & \textbf{Tb} \\
\hline
Load & 2.77 & 7.86 & 24.71 & 372.24 \\
Normalize & 4.47 & 13.73 & 117.90 & 89.52 \\
Create trigrams & 9.50 & 142.58 & 378.16 & 861.33 \\
Create words & 0.38 & 1.83 & 9.03 & 15.80 \\
Parse event IDs & 4.69 & 137.18 & 273.67 & 716.53 \\
\hline
\end{tabularx}

\end{table}

\section{Discussion}
\textbf{RQ1} regarded the model performance in terms of AUC-ROC, F1-score and time. As the results indicate, OOVD reached the best overall AUC-ROC and F1-scores despite failing on HDFS. Time-wise OOVD also performed the best with RM as a close second. In general, the times were better than suggested by previous work \cite{expreport}. For \textbf{RQ2}, we examined the log representation and found that trigram-representation can reach the best results but it is computationally much more expensive than the word-representation. Finally, for \textbf{RQ3}, we studied the impact of filtering out anomalies from the training data, Unfiltered vs. Normal-only scenario. Normal-only scenario improved the overall results except for the IF model and the event-representation. In fact, IF model with events had the best average results for Unfiltered training data and OOVD with trigrams for Normal-only data.

One of the key contributions of this study is the insight on the characteristics of common benchmark datasets. For example, the performance of the OOVD tells how reliant the data is on being able to detect new items, or in other words, perform novelty detection. The high scores on BGL and Tb with the OOVD essentially suggest that novelty detection is all you need with these event-level supercomputer logs. Conversely, on HDFS the other models performed much better, so there novelty detection has little value. In fact, for HDFS you only need to find the rare instances of words and the sequence is likely to be anomalous making RM the best choice. 

\begin{figure}
    \centering
    \includegraphics[width=0.7\columnwidth]{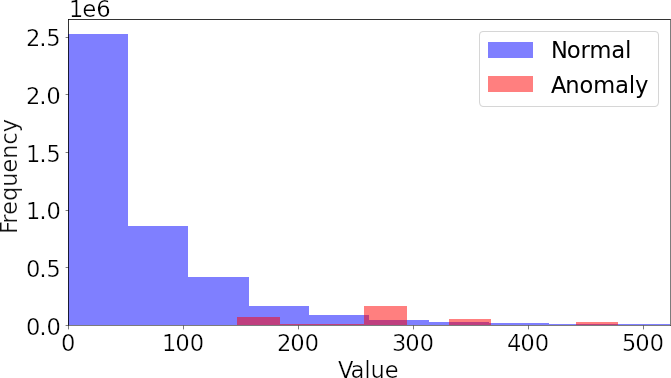}
    \caption{Distribution of the RM scores on BGL (Word-representation, filtered setup)}
    \label{fig:bglwordsdist}
\end{figure}

By emphasizing the AUC-ROC metric this study sheds light on methods that show potential but might have difficulty setting a threshold for anomaly detection. Fig.~\ref{fig:bglwordsdist} illustrates the distribution of RM scores for BGL using the word-representation. What the figure shows is that if we could assign a single value which would split the blue and red areas completely, we would have a perfect classifier.  Of course in a real use case you would not see the colours. We can use such figures for aiding threshold selection.  

Now, consider visualizing both red and blue bars as black in Fig.~\ref{fig:bglwordsdist} and observe the surprising gap-up in the otherwise downward occurring near 250. This gap-up represents an ideal point for setting a threshold.  If we set the threshold at 250 as suggested by the figure we would get an accuracy of 0.957 and an F1-score (binary) of 0.721. If we can automate this process, we could perform unsupervised threshold selection as well. However, due to the complexity of such process, it was out of scope of this paper, but it remains a possible topic for future research. 

While the F1-score was considered as a secondary metric in this study, it provides a better connection to the existing literature such as Loglizer \cite{expreport} and AutoLog \cite{autolog} than only using AUC-ROC. Direct comparisons are still infeasible due to differences in preprocessing and label usage, but results show similarities, for example in the poor performance of the IF model on certain test cases. Further, this paper aims to contribute by presenting various models and representation methods for researchers that encounter low performance (as measured by F1-score) on dataset specific cases. Also, in metric selection itself, this study showcased that there is only a moderate correlation (0.43) between F1-score and AUCROC across all test cases. In essence, this means that generalizing the scores across datasets is difficult and calls for appropriate attention to metric selection.  

While the Normal-only test scenario gained better results than the Unfiltered, it should be noted that there is a fundamental difference with these two approaches in what is actually being examined. In the Unfiltered approach, one is completely at the mercy of the data, and if there are a large number of anomalies, it is impossible to identify which are anomalies and which are not. The Unfiltered approach essentially relies on the assumption that rarity equates anomalousness. To some extent, the models can group similar rows under the same category which can cause a rare individual row to be correctly interpreted as normal, making the model potentially better than a simple frequency table with a threshold. Whereas, with the Normal-only data filtering, we do not need to equate rarity with anomalousness, but instead, the analysis is based on the deviation from normal. Then, the fundamental question is not what is rare enough but, rather, what is different enough. As such, we argue that in unsupervised learning, training with only normal data better suits anomaly detection while the unfiltered approach is closer to outlier detection. For academics, these results emphasize the importance of communicating the data selection process clearly. 

This study opens various directions for future research. For example, finding a way to combine OOVD with other models in a way that they do not clash with each others' results has the promise to reach top scores on each dataset. Another direction to extend these results is to introduce new configurations for the data preprocessing, such as only using chronologically earlier data to predict later events as was done in a recent semi-supervised study \cite{lglog}. The premise of this paper, fast and unsupervised anomaly detection, in itself encourages further studies to move toward more practical application of the anomaly detection approaches. 

As a limitation of the study, parameters are very crucial on unsupervised methods but systematic parameter optimization for the models was not in the scope of this paper. Additionally, utilizing the labels for the F1-score is not realistic and not necessarily comparable to those of similar previous studies \cite{expreport}, \cite{loganomaly}. We found that the results were very stable on BGL, HDFS and Thunderbird, which warranted us to only report the results of just one run. For Hadoop, though, the results varied quite a bit. This is a limitation of the study that can be addressed in future research. 

\section{Conclusion}
This paper investigates speed and performance of anomaly detection using unlabeled data. Our findings indicate that our simple custom models (OOVD and Rarity Model) outperform established models (KMeans and IF) in terms of speed. In terms of performance (AUC-ROC, F1-score), IF emerges as a viable option when dealing with Unfiltered training data comprising both normal instances and anomalies. OOVD appears to be the preferred choice when the training data is Normal-only. However, caution is advised since even the top-performing models exhibit subpar performance in certain scenarios. Nevertheless, these insights offer a good starting point for those aiming to conduct fast, unsupervised anomaly detection.

\section*{Acknowledgment}
Funded by the Academy of Finland (grant ID 349487).

\bibliography{ref}
\bibliographystyle{IEEEtranN}

\end{document}